\documentclass{article}

\usepackage[english]{babel}

\usepackage[letterpaper,top=2cm,bottom=2cm,left=3cm,right=3cm,marginparwidth=1.75cm]{geometry}

\usepackage{amsmath}
\usepackage{graphicx}
\usepackage{subcaption}
\usepackage{authblk}
\usepackage{multirow}
\usepackage[colorlinks=true, allcolors=blue]{hyperref}

\providecommand{\keywords}[1]
{
  \small	
  \textbf{\textit{Keywords---}} #1
}

\title{\textbf{Robust control of Z-source inverter operated BLDC motor using Sliding Mode Control for Electric Vehicle applications}}

\author{Gourab Das\textit{(Student Member, IEEE)}\thanks{Email: gourab@ieee.org}}
\author{Dibakar Das\thanks{Email: ddasdibakardas@gmail.com}}
\author{Md Arif\thanks{Email: mdarif11099@gmail.com}}
\author{Biplab Satpati\thanks{\textbf{Corresponding Author Email: biplab.satpati@gmail.com}}}
\affil{Department of Electrical Engineering\\ University Institute of Technology, The University of Burdwan}
\begin{document}
\maketitle

\begin{abstract}
The rapid development and expansion of the EV market marked by the advent of third decade of the 21st century has improved the possibility of a sustainable automotive future. The present EV drivetrain run by BLDC motor has become increasingly complicated thus requiring efficient and accurate controls. The paper begins with discussing the problems in existing models, the research then focuses on increasing the robustness of the system towards disturbances and uncertainties by using Sliding Mode Control to control the ZSI, which has been chosen as the main power converter topology in place of VSI or CSI. The introduction of SMC has improved the performance of the drivetrain when applied with Vehicle dynamics over a Drive Cycle.
\end{abstract}

\section{Introduction}
The popularity of electric vehicles (EVs) is increasing due to their environmental benefits, energy efficiency, and low operating costs. However, the performance of an EV is dependent on its drive train system and control strategy. Brushless DC (BLDC) motors are widely used in EVs due to their high efficiency, longer working duration and life as well as low sustention, but they have some limitations, including the need for a rotor position sensor and complex control algorithms \cite{nian2014regenerative}.
\par To address these issues, research has been conducted on developing a robust controller for EVs with a modified converter that utilises sensorless control algorithms to predict the rotor position. This modification eliminates the need for a rotor position sensor, reduces complexity and cost, and provides improved performance and reliability. Contemporary control techniques have been used such as sliding mode control and adaptive control to enhance the safety and reliability of the EV under different driving conditions.
\par The proposed research aims to design and develop a robust controller for EVs with a modified converter, evaluate its performance through simulations and experimental studies, and conduct a comparative analysis with existing controller for EVs. The outcomes of this research will contribute to the development of efficient and reliable EVs, which are crucial for achieving sustainable transportation \cite{tashakori2013fault}.

\section{Literature Review}
Electric vehicles are gaining popularity as an eco-friendly and sustainable mode of transportation, and researchers have been continuously striding towards their advancement. Converter is one of the most essential components in an electric vehicle and it is responsible for conversion of DC power to AC power for the motor. While existing converter technologies works well in most cases, there are several drawbacks that can limit the overall performance of the electric vehicle \cite{tashakori2011characteristics}.
\par One of the major issues with the existing converter technology is its limited ability to handle dynamic loads. In electric vehicles, the load on the motor can vary significantly, depending on dynamic factors such as road gradient, vehicle speed, and acceleration. The existing converter topology which mostly uses Voltage Source Inverter (VSI) has limited  ability of handling these dynamic loads efficiently, resulting in reduced efficiency of the electric vehicle, similar problems are faced by Current Source Inverters (CSI). VSI also has low efficiency at partial loads, along with its limited ability to control motor torque and speed it requires complex control algorithms and hardware. These limitations can result in energy loss, reduced driving range, and increased costs, therefore fuelling the need for  alternative power electronic topologies \cite{ehsani2007hybrid}.
\par Another major concern with the existing converter technology is its sensitivity to parameter variations. Most of the research has developed the control schemes on the grounds of regular industrial control i.e. PID. Upon reviewing existing research papers on vehicle dynamics and drive cycle, it was observed that most studies do not combine both aspects and therefore do not categorically look into the complete problem of reducing complexity, thus relying solely on PID control, and in some cases, using either the Runge-Kutta or Zeigler-Nicholas method to train the model. While these methods may yield satisfactory results, they do not take into account the complexities and uncertainties associated with real-world driving scenarios. Although PID or similar control schemes reduce a lot of complexity and can superlative perform for a single type of terrain, however, when we consider the vehicle dynamics and drive cycle to simulate real world complexities  such as wind, road conditions, and other external factors and thereafter apply PID control, the performance is often less than what is desired. This suggests that there is a need to develop new algorithms or methods that can address the challenges of randomness in driving and can be used with advanced controllers, which are more effective than traditional industrial controllers in terms of robustness \cite{tashakori2012direct}.
\par SMC is a method that have been used to address the issues and has been used in the scope of present research but the use of High-Order Sliding Mode Control (HOSMC) that provides better tracking performance and reduced chattering\cite{liu2014observer} but the complexity of computation resulting in delayed response makes the real time applicability harder than in lower order systems. Another example is the use of adaptive SMC for speed control in EV drive trains. Adaptive SMC adjusts the control parameters based on the system dynamics to improve tracking performance and reduce chattering \cite{jang2002position}, the concern being that the fine tuning is required to employ it in physical systems which has unpredictable disturbances thus are humongous and can be managed by only expensive and advanced sensing mechanism.

\section{Research Methodology}
\subsection{Overview}
The challenges and deficits mentioned previously have motivated the present research, the present proposal uses existing topologies and control schemes in a pristine way to design the SIMULINK model so as to improve the modeled vehicle efficiency and robustness. The new model replaces the most frequently used VSI with a ZSI in the power distribution system which outperforms the former due to higher voltage conversion ratios and better harmonic performance.
\par After conducting an extensive review of research methodologies in current studies, we have implemented a new approach to enhance the efficiency and robustness of the SIMULINK model.

\subsection{Battery}
Battery EVs or BEVs and their accessory systems rely exclusively on batteries, various batteries have been contesting the automotive field Lead-Acid (Pb-A), Nickel-Cadmium (Ni-Cd), NickeL-metal-Hydride (Ni–MH), Lithium-ion (Li-ion), Lithium-ion Polymer (Li-polymer) and Sodium Nickel Chloride (NaNiCl) but Li-ion has gradually outlived all competitive threats from all other batteries on various factors like transmission efficiency, power delivery requirement and cost constraint including Li-polymer batteries which came into existence after Li-ion batteries \cite{manzetti2015electric}. 
\subsubsection{Performance and Environmental Impact}
Li-ion batteries offer several advantages over other types of batteries, such as higher energy density which results in a lighter and more compact battery, longer lifespan leading to less frequent battery replacements, and shorter charging time, ultimately improving convenience and additionally has lesser toxicity. The biggest benefit of the Li-ion battery is its self-discharge rate so charge is stored for a long duration of time when not in use, which is very important from a consumer perspective \cite{ding2019automotive}. Li-ion has specific shortcomings such as higher cost and limited availability but the benefits outlive the constraints. We have extensively used Li-ion battery models as it has industry relevance \cite{peters2017environmental}.

\subsubsection{Battery Management System (BMS) and Charging}
Modern BMS have evolved to include multiple functionalities such as monitoring the state of health and state of power, optimising battery performance, and other parameters such as ageing and residual factors and handling cooling of the battery model to ensure the battery pack's lifespan extension, and operation at optimal conditions and is safe for operation. The BMS optimises the EV's range and performance by managing the dissipation of individual cells connected in series, parallel, or hybrid  \cite{liu2019brief}.
\par Charging circuit is essential in providing power to the battery which in turn powers the EV; various chargers and charging strategies like Luenberger observer, Kalman Filter and sliding-observer etc. \cite{barillas2015comparative} have been implemented for various types of battery charging, most common of them are level-1 charging, level-2 charging, high voltage DC charging and wireless charging using AC-DC converters, DC-DC converters or Bi-directional converters \cite{sujitha2017res}. Incorporation of charging capabilities could be added to the model which presently is not in our scope, but this does not compromise the models’ accuracy.\\ \\
\textbf{Mathematical Model: }
The generalised discharge model $(\hat{i} >0)$ of a battery can be given by:
\begin{equation}
f(i \cdot t,\hat{i},i) = E_0 -\kappa_b \frac{Q}{Q- i\cdot t} \cdot \hat{i}- \kappa_b \frac{Q}{Q- i\cdot t}+A e^{-B \cdot i \cdot t}
\end{equation}
The generalised charging model  $(\hat{i} <0)$ can be given by:
\begin{equation}
f(i \cdot t,\hat{i},i) = E_0 -\kappa_b \frac{Q}{ i\cdot t + 0.1\, Q} \cdot \hat{i}- \kappa_b \frac{Q}{Q- i\cdot t}+A e^{-B \cdot i \cdot t}
\end{equation}
\par where $E_0$ represents constant voltage , $\kappa_b$ represents the polarisation constant, $\hat{i}$ represents current dynamics in low frequency , $Q$ represents the higest battery capacity, $A$ represents exponential voltage, $B$ represents the exponential capacity, and $i\cdot t$ represents extractable capacity\cite{thakkar2021electrical}.

\subsection{Z-source Inverter}
The ZSI network was introduced in 2002 as a series of passive components, such as inductors and capacitors, connected in a unique single stage configuration allowing the inverter to provide voltage boost, voltage buck, and voltage inversion functions making it a versatile power conversion device and was named as Z-source inverter \cite{liu2016impedance}. Today ZSI is used for AC-AC, DC-DC, AC-DC, DC-AC type conversion \cite{siwakoti2014impedance}, the invention of the PWM scheme for ZSI \cite{loh2004pulse} improved its reach.
\subsubsection{Performance}
ZSIs have a higher voltage boost capability compared to VSIs and CSIs, and has wider range of input voltages. Secondly, ZSIs can operate in both buck and boost modes simultaneously, so a higher power can be delivered at low DC input. Thirdly, ZSIs are less susceptible to electromagnetic interference (EMI) improving reliability. Finally, ZSIs have a simpler topology which makes them cheaper and easier to design, implement, and control. 
\par Overall, they are an attractive option in renewable energy systems in Maximum Power Point Tracking, in motor drives where broader ranges of output voltage and current is essential, and in Uninterrupted Power Supply devices for power quality improvement, etc. \cite{raveendhra2022effects}.
\subsubsection{Operating Principle}
The operating principle of a ZSI is modeled by the concept of impedance-source networks to regulate the DC input voltage, which is made of a pair of series inductors and parallel capacitors forming a "shoot-through" impedance network. Upon switching on, the impedance network initially acts as a current source, charging the capacitor, upon further operation, it acts as a voltage source, boosting the input voltage, delivering high output power \cite{liu2012permanent}. The shoot-through capability enables the ZSI to operate both in buck and boost modes simultaneously, requiring a smaller DC input voltage while delivering higher output power providing it a higher voltage boost capability and a wider range of operating voltages compared to traditional inverters.\\ \\
\textbf{Mathematical Model: } The inductors and capacitors are identical so we consider $L1=L2=L$ and $C1=C2=C$. The basic operating equations required for ZSI are :
\begin{equation}
\begin{aligned}
V_{L} &= L(\frac{di}{dt}) \\
V_{C} &= \frac{1}{C} \int i dt \\
V_{o} &= (1 + \Delta)\,V_{s'} - \Delta V_{L'} \\
i_{o} &= (1 - \Delta)\,i_{s'} + \Delta\,i_{L'} \\
B &= \frac{1}{1-2\Delta} \geq 1
\end{aligned}
\end{equation}
\par where $V_{L'}$ and $i_{L'}$ are the inductor voltage and current, $V_{C}$ is the capacitor voltage, $V_{s'}$, $i_{s'}$, $V_{o}$, $i_{o}$ are the input and output voltage and current respectively and $B$ is the boost factor of the inverter. $\Delta$ represents the duty cycle of the pulse width modulation (PWM) which is given by the percentage of time during which both upper and lower switches are turned on, which determines the amount of time the input voltage is switched to the output and primarily determines the on-off frequency \cite{pilehvar2015inverter}.
\subsubsection{Control Schemes}
The most widely used concept is PWM control which performs the control action by switching off the power devices by modulating the shoot-through duty cycle ($\Delta$), to achieve a desired output waveform in terms of current and voltage \cite{liu2012permanent}.
\par The two main types of PWM control techniques used for ZSI are traditional and State vector based \cite{mande2020comprehensive}, in state vector based control the operation is done by dividing the voltage and current into three vectors and in traditional it is done by comparison of various signals; their choice is done on the basis of requirement.
\par In recent times, a fuzzy logic, Maximum Power Point Tracking control and Predictive Control approach have also been adopted as they offer better performance in nonlinear systems or systems which have various degrees of uncertainty in the system \cite{singh2019comprehensive}.

\subsection{Brushless DC (BLDC) motor}
A BLDC motor is a category of permanent magnet synchronous motor which uses electronic controllers alternative to brushes for controlling the current flow through the stator windings, which in turn produces a magnetic field that energises the permanent magnets on the rotor to produce torque. The commutation system of a BLDC motor typically consists of position sensors, such as Hall effect sensors or optical encoders, and an electronic controller \cite{yedamale2003brushless}; in advanced models sensorless working is also seen. New control techniques and power electronics have expanded the potential applications of BLDC motors \cite{tibor2011modeling}.
\subsubsection{Operating Principle and Optimisation}
BLDC motor constitutes power electronic converters, permanent magnet-synchronous machine (PMSM), sensors, and control systems \cite{baldursson2005bldc}. The operation of a brushless motor is typically achieved through a three-phase bridge inverter; to ensure suitable commutation in every $60^{\circ}$, a rotor position sensor is required. The electronic commutation eliminates issues like sparking and wearing out, providing greater durability compared to traditional counterparts. Sensorless control of BLDC motors is a method that estimates rotor position using information about the motor's voltage, current, and back-emf instead of using physical sensors further reducing costs. 
\par The conversion of mechanical to electrical energy is done by the PMSM, while the positional sensors feed inputs to monitor algorithms to apprehend gate pulses to power semiconductor switches of the converter have its origin on the rotor position and reference signals like torque, voltage, speed, etc. \cite{jeon2000new} regulating the overall operation. In general, two types of drivers can power PMSMs which generate either sinusoidal or non-sinusoidal back electromotive force (EMF) waveforms: those based on voltage sources and those based on current sources. Machines with sinusoidal back EMF can maintain almost constant torque, whereas those with non-sinusoidal back EMF can be operated using smaller inverters with lower losses for the same power level.
\par Optimising a BLDC motor involves enhancing the accuracy, stability, and responsiveness of the motor's speed or torque control using PID, Field oriented Control and Sliding Mode Control which has been applied in detail in the present scope of research. Various optimization techniques such as Genetic Algorithm, Particle Swarm Optimisation, etc. can be used to find the optimal gain values \cite{shao2006improved}.
\\ \\
\textbf{Mathematical Model: }The PMSM motor is composed of a rotor with permanent magnets and three stator windings. The rotor and stainless steel have high resistivity, hence the rotor-induced currents can be neglected. The voltage equation is formulated in phase variables but the damper windings are unaccounted.
\begin{equation}
    \begin{bmatrix}
    V_{as}\\ V_{bs} \\ V_{cs}
    \end{bmatrix}
    =
    \begin{bmatrix}
    R_{sp} & 0 & 0 \\ 0 & R_{sp} & 0 \\ 0 & 0 & R_{sp}
    \end{bmatrix}
    \begin{bmatrix}
    i_{ap} \\ i_{bp} \\ i_{cp}
    \end{bmatrix}
    + \frac{d}{dt}
    \begin{bmatrix}
    L_{aa} & L_{ab} & L_{ac} \\ L_{ba} & L_{bb} & L_{bc} \\ L_{ca} & L_{cb} & L_{cc}
    \end{bmatrix}
    \begin{bmatrix}
    i_{ap} \\ i_{bp} \\ i_{cp}
    \end{bmatrix}+
    \begin{bmatrix}
    e_{ap}\\ e_{bp} \\ e_{cp}
    \end{bmatrix}
\end{equation}
The stator phase voltages are represented by $V_{as}$, $V_{bs}$, and $V_{cs}$. The stator resistance per phase is denoted by $R_{sp}$. The stator phase currents are represented by $i_{ap}$, $i_{bp}$, and $i_{cp}$. The self-inductance of phases $a$, $b$, and $c$ are denoted by $L_{aa}$, $L_{bb}$, and $L_{cc}$, respectively. The mutual inductances of phases $a$, $b$, and $c$ are shown by $L_{ab}$, $L_{bc}$, and $L_{ca}$. The back electromotive forces for individual phases are denoted by $e_{ap}$, $e_{bp}$, and $e_{cp}$. Assuming uniform winding resistance and a non-salient rotor, it is considered that there is no variation in rotor reluctance with angle. \cite{park2000new}. Which means that $L_{aa} = L_{bb}= L_{cc}= L_s$ and also $L_{ab} = L_{ba}= L_{bc}=L_{cb} = L_{ca}= L_{ac}= L_m$.
The modified equation can be given as:
\begin{equation}
    \begin{bmatrix}
    V_{as}\\ V_{bs} \\ V_{cs}
    \end{bmatrix}
    =
    \begin{bmatrix}
    R_{sp} & 0 & 0 \\ 0 & R_{sp} & 0 \\ 0 & 0 & R_{sp}
    \end{bmatrix}
    \begin{bmatrix}
    i_{ap} \\ i_{bp} \\ i_{cp}
    \end{bmatrix}
    + \frac{d}{dt}
    \begin{bmatrix}
    L_s & L_m & L_m \\ L_m & L_s & L_m \\ L_m & L_m & L_s
    \end{bmatrix}
    \begin{bmatrix}
    i_{ap} \\ i_{bp} \\ i_{cp}
    \end{bmatrix}+
    \begin{bmatrix}
    e_{ap}\\ e_{bp} \\ e_{cp}
    \end{bmatrix}
\end{equation}
where, the phase voltages $V_{as}$, $V_{bs}$, and $V_{cs}$ are 
$V_{as}=V_{ao}-V_{no},V_{bs}=V_{bo}-V_{no},V_{cs}=V_{co}-V_{no}$ and $V_{ao}$, $V_{bo}$, $V_{co}$ and $V_{no}$ are three phase and neutral voltages respectively.
Since stator phase currents have to be equipoised i.e. 
\begin{equation}\label{bldcref0}
    i_{ap}+i_{bp}+i_{cp}=0
\end{equation}
thus the matrix of inductances in the models can be given by,
\begin{equation}
    L_m i_{ap}+L_m i_{bp}=-L_m i_{cp}
\end{equation}
So we get the state space from the final equation as:
\begin{equation}\label{bldcref1}
    \begin{bmatrix}
    V_{as}\\ V_{bs} \\ V_{cs}
    \end{bmatrix}
    =
    \begin{bmatrix}
    R_{sp} & 0 & 0 \\ 0 & R_{sp} & 0 \\ 0 & 0 & R_{sp}
    \end{bmatrix}
    \begin{bmatrix}
    i_{ap} \\ i_{bp} \\ i_{cp}
    \end{bmatrix}
    + \frac{d}{dt}
    \begin{bmatrix}
    L_s-L_m & 0 & 0 \\ 0 & L_s-L_m & 0 \\ 0 & 0 & L_s-L_m
    \end{bmatrix}
    \begin{bmatrix}
    i_{ap} \\ i_{bp} \\ i_{cp}
    \end{bmatrix}+
    \begin{bmatrix}
    e_{ap}\\ e_{bp} \\ e_{cp}
    \end{bmatrix}
\end{equation}
It has been feigned that back EMF $e_{ap}$,  $e_{bp}$, and $e_{cp}$ are in trapezoidal nature so
\begin{equation}
    \begin{bmatrix}
    e_{ap} \\ e_{bp} \\ e_{cp}
    \end{bmatrix}
    =
    \omega_m \lambda_m
    \begin{bmatrix}
    F_{a} (\theta_r) \\ F_{b} (\theta_r) \\ F_{c} (\theta_r)
    \end{bmatrix}
\end{equation}
where $\omega_m$ the angular speed of the rotor in radians per seconds, $\lambda_m$ is the amount of flux linkage, $\theta_r$ is the position of rotor in radian and the functions $F_{a}(\theta_r)$, $F_{b}(\theta_r)$ and $F_{c}(\theta_r)$, have the same shape as  $e_{ap}$, $e_{bp}$, and $e_{cp}$ \cite{hwang2012design, choi2001global}. \\
The electromagnetic torques are obtained as the derivatives of the flux linkages, which are continuous functions \cite{shifat2020effective}. The flux density function is also smoothed out due to fringing, resulting in no sharp edges. \\
We know that the electromagnetic torque can be defined in Newtons as $T_e = [e_{ap} i_{ap} + e_{bp} i_{bp} +e_{cp} i_{cp}]/\omega_m$.\\ We know that moment of inertia can be defined as $J=J_m+J_l$
The equation for a basic motion system includes variables such as inertia ($J$), friction coefficient ($B$), and load torque ($T_l$) is the following:
\begin{equation}\label{bldcref2}
    J \frac{d\omega_m}{dt}+B\omega_m=(T_e+T_l)
\end{equation}
Also we know that:
\begin{equation}\label{bldcref3}
    \frac{d\theta_r}{dt}=\frac{p}{2}\,\omega_m
\end{equation}
The damping coefficient B is usually small and can be neglected in the system. The equation mentioned above describes the rotor position $\theta r$, which repeats every $2\pi$. In order to avoid an imbalance in the applied voltage and simulate the performance of the drive, it is necessary to consider the potential of the neutral point with respect to the zero potential $V_{no}$. From the volt-ampere equation (\ref{bldcref1}) we can get,
\begin{equation}\label{bldcref4}
        V_{ao}+V_{bo}+V_{co}-3V_{no} = R_{sp}(i_{ap}+i_{bp}+i_{cp})+(L_s-L_m)(pi_{ap}+p{ib}_p+pi_{cp})+(e_{ap}+e_{bp}+e_{cp})
\end{equation}
Equating (\ref{bldcref0}) with (\ref{bldcref4}) we get:
\begin{equation}
    \begin{aligned}
        V_{ao}+V_{bo}+V_{co}-3V_{no} &= (e_{ap}+e_{bp}+e_{cp})\\
        V_{no} &= [(V_{ao}+V_{bo}+V_{co})-(e_{ap}+e_{bp}+e_{cp})]/3
    \end{aligned}
\end{equation}
The assortment of differential equations alluded in equations (\ref{bldcref1}), (\ref{bldcref2}), and (\ref{bldcref3}), establishes the exhibited model in terms of the variables $i_{ap}$, $i_{bp}$, $i_{cp}$ , $\omega_m$ and, $\theta_r$ time as an independent variable.
Assembling all the incidental equations, the system equations in state-space form are:
\begin{equation}
    \dot{x}=Ax+Bu+Ce
\end{equation}
where,
\begin{equation}
    \begin{aligned}
        x &= {\begin{bmatrix}i_{ap} & i_{bp} & i_{cp} & \omega_m & \theta_r \end{bmatrix}}^T\\
        A &= 
        \begin{bmatrix}
        -\frac{R_{sp}}{L_s-L_m} & 0 & 0 & -\frac{\lambda_m}{j} F_{a}(\theta_r) & 0 \\
        0 & -\frac{R_{sp}}{L_s-L_m} & 0 & -\frac{\lambda_m}{j} F_{b}(\theta_r) & 0 \\
        0 & 0 & -\frac{R_{sp}}{L_s-L_m} & -\frac{\lambda_m}{j} F_{c}(\theta_r) & 0 \\
        -\frac{\lambda_m}{j} F_{a}(\theta_r) & -\frac{\lambda_m}{j} F_{b}(\theta_r) & -\frac{\lambda_m}{j} F_{c}(\theta_r) & -\frac{B}{J} & 0 \\
        0 & 0 & 0 & \frac{P}{2} & 0 
        \end{bmatrix}\\
        B &=
        \begin{bmatrix}
        \frac{1}{L_s-L_m} & 0 & 0 & 0\\
        0 & \frac{1}{L_s-L_m} & 0 & 0\\
        0 & 0 & \frac{1}{L_s-L_m} & 0\\
        0 & 0 & 0 & \frac{1}{L_s-L_m}
        \end{bmatrix}\\
        C &=
        \begin{bmatrix}
        \frac{1}{L_s-L_m} & 0 & 0 \\
        0 & \frac{1}{L_s-L_m} & 0 \\
        0 & 0 & \frac{1}{L_s-L_m} 
        \end{bmatrix}\\
        u &= {\begin{bmatrix}V_{as} & V_{bs} & V_{cs} & T_l \end{bmatrix}}^T \\
        e &= {\begin{bmatrix} e_{ap} & e_{bp} & e_{cp} \end{bmatrix}}^T
    \end{aligned}
\end{equation}
\subsubsection{Application}
BLDC motors present certain advantages over conventional DC motors, including higher efficiency and power density, longer lifespan, low maintenance, reliability, and silent operation. Their application range from hard disk drives, DVD and Blu-ray players in consumer electronics and fans, pumps, ventilators etc. in aerospace and medical industry \cite{tashakori2011modeling}.
\subsection{Control System}
\subsubsection{Open Loop Control}
The open-loop control of a BLDC motor is achieved by fixed input voltage to the motor to achieve a pre-programmed voltage profile or speed torque characteristics without measuring the speed or position. Though the method is cheap and simple, it may not provide the precise control needed, making the motor susceptible to disturbances like load or voltage fluctuations. Thus, closed-loop control methods are often preferred for all real world requirements including BLDC motors, where feedback sensors monitor speed and position and adjustments in input voltage are computed \cite{borovic2005open}.
\subsubsection{Proportional-Integral-Derivative (PID) Control}
PID control is a closed-loop control mechanism that computes an error signal between a desired and measured process variable like temperature, pressure, flow rate, after which proportional, integral, and derivative operations are applied to match system output to the reference value and maintain it. The combination provides a stable and accurate control mechanism that can be used in various industries including manufacturing, aerospace, and automotive including BLDC motor control.
\par The PID controller's output can be represented as:
\begin{equation}
u(t) = \kappa_p\,\varepsilon(t) + \kappa_i \int \varepsilon(\tau)\,d\tau + \kappa_d\,\frac{d\,\varepsilon(t)}{dt}
\end{equation}
\par where $u(t)$ signifies the control signal, $\varepsilon(t)$ represents the error signal between desired setpoint or reference and the actual output, and $\kappa_p$, $\kappa_i$, and $\kappa_d$ are the proportional, integral, and derivative gains respectively. These gains are fine-tuned to achieve precised response of the system either through trial and error or optimization \cite{johnson2005pid}.
\subsubsection{Sliding Mode Control (SMC)}
SMC is a nonlinear control technique for dynamic systems introduced by Russian mathematician Emelyanov in the 1950s to improve robustness of the systems with uncertainties and disturbances. In sensorless control of BLDC motor by SMC  a sliding variable is calculated by computing the difference of the estimated to the desired rotor position precisely. Sensorless control eliminates the cost and complexity of sensors, reducing maintenance requirements, making it a cost-effective and reliable alternative to sensor-based control\cite{kaynak2001fusion}.
\par SMC uses a sliding surface to separate a system's states into two regions and guide its state variables to the surface. A feedback control law moves the system along the sliding surface to arrive at the desired state by ensuring the difference between the system's states and the surface reaches the minimum \cite{shtessel2014sliding}. SMC is robust and adaptive in systems with uncertainties and disturbances. The lesser necessity of prior knowledge of system's dynamics and lower complexity in implementation makes it a popular choice for real-world applications.\\ \\
\textbf{Mathematical Explanation of SMC: } Consider an unit mass undergoing single dimensional motion, being pulled by the control input force $u$ and resisted by the bounded disturbance force $f(x_1,x_2,t)$ including the viscous friction force as well as the unknown rigid forces associated with this mass. The system states are designated by $x_1$ and $x_2$, where $x_1$ represents position and $x_2$ represents velocity of the body \cite{young1999control}. The state space equation of the motion of the block can be represented, where the disturbance force is bounded by a positive quantity:
\begin{equation}
\begin{aligned}
\dot{x}_1 &= x_2 \\
\dot{x}_2 &= u + f(x_1, x_2, t)\\
|f(x_1, x_2, t)| &\leq L > 0   
\end{aligned}
\end{equation}
\par where L is any positive real number.\\
\textit{Control Problem:} The objective of the control solution is to design a state feedback control law:
\begin{equation}
    \begin{aligned}
        u &=u (x_1,  x_2) \\
        u &=-kx \\
        \lim\limits_{n\to\infty} (x_1, x_2) &=0
    \end{aligned}
\end{equation}
such that it can asymptotically drive the system states to point of equilibrium, that is the system states reach the equilibrium point as time varies from $0$ to infinity.
Considering the state feedback control law for a two state system,
\begin{equation}
    u = - k_1\,x_1 - k_2\,x_2
\end{equation}
where both $k_1$ and $k_2$ are both greater than zero, and the system stabilises only when $f (x_1,x_2,t)=0$.\\
In real world systems there will be disturbance which will haul the system states to a bounded domain $\delta$. \cite{drakunov1992sliding,levant1993sliding}.
So, $\delta$ is a function of $k1, k2$ and which drives the system states to a bounded domain for a given disturbance. Thus, it will not bring it to convergence but it will bring it close to the stability. That means, it will not converge to $0$ or equilibrium point provided the system disturbances are not equal to $0$.
The overall state dynamics of the system can be represented by a homogeneous differential equation of first order,
\begin{equation}
    \dot{x_1}+ cx_1 = 0 
\end{equation}
where c $>$ 0 and where state variable $x_2(t)$ represents,
\begin{equation}
    x_2(t) = \dot{x_1}(t)
\end{equation}
for this system, we have the solution as 
\begin{equation}
    x_1(t) = x_1(0)e^{-ct}
\end{equation}
And, the derivative of the solution as
\begin{equation}
    x_2(t) = -c x_1(0)  e^{-ct}
\end{equation}
This shows that the system states $x_1, x_2$ converge to  an equilibrium point asymptotically. From these expressions we can form the solution of this differential equation. We have observed that compensated state dynamic equation has no disturbance effect on the system, which means $f(x_1,x_2,t)=0$ the disturbance effect has no consequences on the compensated state dynamic equation.
A system dynamics variable $\sigma$ is introduced where $\sigma$ is a function of $x_1$ and $x_2$ i.e. $\sigma(x_1,x_2)$ and can be represented by 
\begin{equation}
    \sigma(x_1,x_2) = x_2 + c\, x_1
\end{equation}
where c is any positive number. To achieve asymptotic convergence, the state variables $x_1$ and $x_2$, with the effect of disturbance in effect there is a need for a sliding mode control law to drive the system to asymptotic stability by converging the variable $\sigma$ in finite time. Lyapunov function techniques used to analyse the dynamics as in  \cite{liu2011advanced}
Applying Lyapunov function to analyse stability and to be of use in controller design, 
\begin{equation}
  \begin{aligned}
      V &={\frac{1}{2}} \sigma^2\\
      V &= {\frac{1}{2}}{(x_2 + cx_1)}^2
  \end{aligned}
\end{equation}
where the system asymptotically converges only when:
\begin{itemize}
    \item $V$ must be positive definite and $V \leq 0$
    \item $\lim\limits_{\sigma\to\infty} V = \infty$
\end{itemize}
For achieving finite time convergence we need to modify $V \leq 0$ as
\begin{equation}
    \dot{V} \leq -\alpha V^{\frac{1}{2}}
\end{equation}
where $\alpha$ is a positive constant. So we have derivation as
\begin{equation}
    \frac{1}{\alpha} V^{-\frac{1}{2}} dv \leq - \, dt
\end{equation}
Integrating both sides we can get,
\begin{equation}
    \frac{1}{\alpha}  \int_{V(0)}^{0} v^{-\frac{1}{2}} \,dv \leq - \int_{0}^{t_f} dt
\end{equation}
Simplifying the equation we get,
\begin{equation}
    t_f \leq {\frac{2}{\alpha}} V(0)^{\frac{1}{2}}
\end{equation}
where $\alpha > 0$ . So, we see that $t_f$ is not infinity and $V$ achieves finite time convergence \cite{ackermann1998sliding}. \\
\textit{Controller Design:} The controller design involves driving the variable $\sigma$ to $0$ before infinite time. The dynamics of $\sigma$ must include the control law $u$, given by
\begin{equation}
\begin{aligned}
   \dot{\sigma} &= \dot{x}_2 + c\dot{x}_1 \\
    \text{which implies,\quad\quad} u &= f(x_1,x_2,t) + cx_2 \\
    \text{we know that,\quad\quad} V &= \sigma\dot{\sigma} \\
    \text{therefore,\quad\quad} \dot{V} &= \sigma(u + f(x_1,x_2,t)) + cx_2 \\
    \text{and,\quad\quad} u &= -cx_2 + v 
\end{aligned}
\end{equation}
Thus,
\begin{equation}
    \dot{V} = \sigma(f(x_1,x_2,t) + v)
\end{equation}
where $v$ is a new variable. Since $f(x_1,x_2,t) \leq L > 0$, which is a positive value, we have $\dot{v} \leq \sigma L + \sigma v$. We select the new variable,
\begin{equation}
\begin{aligned}
    v &= -\rho\text{ sign}(\sigma) \\
    \text{we have,\quad\quad}\dot{v} &\leq |\sigma|L - |\sigma|\rho
\end{aligned}   
\end{equation}
Thus,
\begin{equation} \label{smcref1}
\dot{v} \leq |\sigma|(L - \rho) 
\end{equation}\\
Now, considering $V = \frac{1}{2}\sigma^2$ and $\dot{V} \leq -\alpha V^{1/2}$, we get from the value of $V$ that
\begin{equation}
\begin{aligned}
    |\sigma| &= \pm\sqrt{2V} \\
    \text{so,\quad\quad}V^{1/2} &= |\sigma|
\end{aligned}
\end{equation}
Thus,
\begin{equation}\label{smcref2}
    \dot{V} \leq -\, \alpha \frac{|\sigma|}{\sqrt{2}}
\end{equation}
Equating equations (\ref{smcref1}) and (\ref{smcref2}), we get
\begin{equation}  
-\alpha\,|\sigma|^2 = |\sigma|(L - \rho)
\end{equation}
Therefore, the control gain $\rho$ is given by
\begin{equation}
\rho = L + \frac{|\sigma|}{\sqrt{2}}
\end{equation}
which is the gain of the discontinuous control part. As $L$ and $\alpha$ are constant, $\rho$ will also be a constant. So, the first term of the equation is responsible for the disturbance and the second term is responsible for finite time convergence.
Therefore, the final control law can be represented by
\begin{equation}
u = - c\,x_2 - \rho \, \text{sgn}(\sigma)
\end{equation}

\subsection{Vehicle Dynamics}
Vehicle dynamics refers to the study of how vehicles and their systems such as powertrain, battery system, and overall vehicle structure interact with external factors such as road conditions, air drag and driver inputs. Vehicle dynamics is a critical aspect of EV design and engineering because it impacts the vehicle's performance, efficiency, and safety \cite{jerrelind2021contributions}. Additionally, advanced control systems, such as regenerative braking, torque vectoring, and traction control, have remarkable contributions in optimising the vehicle dynamics of electric vehicles. A deep understanding of vehicle dynamics is essential to maximise the benefits of electric propulsion technology, improve energy efficiency, and enhance driving experience and cut down greenhouse gas emissions \cite{crolla2012impact}.\\
\textbf{Mathematical Model: } The position vector equation of the vehicle can be given by,
\begin{equation}
    \begin{bmatrix} x \\ y \end{bmatrix} = \begin{bmatrix} x_0 \\ y_0 \end{bmatrix} + \begin{bmatrix} cos \theta & -sin \theta  \\ sin \theta & cos \theta  \end{bmatrix}\begin{bmatrix} dx \\ dy \end{bmatrix}
\end{equation}
where $\theta \text{and } dx, dy$ represent the vehicle orientation and the displacement vectors respectively, and the velocity vector equation can be obtained as 
\begin{equation}
    \begin{bmatrix} v \\ \omega \end{bmatrix} =\begin{bmatrix} v_x \\ v_y \end{bmatrix} = \begin{bmatrix} \dot{x} \\ \dot{y} \end{bmatrix} = \begin{bmatrix} cos \theta & sin \theta  \\ -sin \theta & cos \theta  \end{bmatrix}\begin{bmatrix} \dot{dx} \\ \dot{dy} \end{bmatrix}
\end{equation}
The lateral kinetic equation are given by,
\begin{equation}
    \begin{aligned}
        \sum F &= m\,a \\
        \tau &= I \alpha \\
        F_x &= F_{xf}+F_{xr} \\
        F_y &= F_{yf} + F_{yr} \\
        \alpha_f &= \frac{v_y + \omega l_f}{v_x}- \delta_f \\
        \alpha_r &= \frac{v_y - \omega l_r}{v_x}+ \delta_r 
    \end{aligned}
\end{equation}
where, $v_y$ is lateral velocity, $c_v$ is angular velocity, $l_f$ is the separation length between front wheels to the point of centre of mass and similarly $l_r$ for rear wheels and $\delta$ are the steering angles $\alpha$ represents slip angles and $\tau$ represents moment of inertia and $F$ represents the tire forces.
The longitudinal kinetic equations are given by, 
\begin{equation}
    \begin{aligned}
        F_y &= m\,g\,sin \theta \\
        F_r &= \mu m\, g\, cos \theta\\
        F_w &= \frac{1}{2}\,\rho_a\,A_f\,C_D\,v^2  
    \end{aligned}
\end{equation}
where, $F_g\text{, }F_r\text{ and}F_w$ represents climbing force, rolling force and aerodynamic drag respectively, $m$ represents the gross weight of the vehicle, $\theta$ represents the inclination angle, $v$ represents velocity, $C_D$ represents drag coefficient, $r_{wheel}$ represents radius of wheel and the coefficient of friction (CoF) is represented as $\mu$.\\
Therefore the tractive forces are;
\begin{equation}
    \begin{aligned}
        F_{tractive} &= \sum (\text{resistive forces}) + m\frac{dv}{dt}\\
        &= (F_g + F_r + F_w) + m\frac{dv}{dt}
    \end{aligned}
\end{equation}
thus the tractive power is,
\begin{equation}
    P_{tractive} = F_{tractive} \times v
\end{equation}
and the torque is, 
\begin{equation}
    \tau = F_{tractive} \times r_{wheel}
\end{equation}
and thus we can find the tractive energy as,
\begin{equation}
    E_{tractive} = \int_0^T P_{tractive}
\end{equation}
The acceleration resistance can be represented as, 
\begin{equation}
    F_a = \left(m+\frac{\sum J_{rot}}{r^2_{wheel}}\right)\frac{dv}{dt}
\end{equation}
where $F_a$ is accelerating force and $J_{rot}$ is the rotational inertia. Now, 
\begin{equation}
    (F_{trf}+F_{trb})-(F_{rf}+F_{rb}+F_w+F_g) =\lambda m \frac{dv}{dt}
\end{equation}
where $\lambda$ is the rotational inertia constant and $F_{rf} \,\&\, F_{rr}$ are the rolling resistance of the tire. Thus the maximum tractive effort for the front wheel is, 
\begin{equation}
    \begin{aligned}
        w_f &= \frac{1}{L}\left[mg L_b\,cos\theta -\left(\tau_{rf}+\tau_{rb}+(F_w+F_g)\times hg+\lambda m \frac{dv}{dt}\times hg\right)\right]\\
        &= \frac{L_b}{L}mg \,cos\theta -\frac{hg}{L}\left[\left(\frac{\mu mgr\,cos\theta}{hg}+F_w+F_g+\lambda m \frac{dv}{dt}\right)\right]\\
        &= \frac{L_b}{L}mg \,cos\theta -\frac{hg}{L}\left[F_f+F_r\left(1- \frac{r_{wheel}}{hg}\right)\right]
    \end{aligned}
\end{equation}
where $h$ is the height. And the same for rear wheel is,
\begin{equation}
    \begin{aligned}
        w_r &= \frac{1}{L}\left[mg L_a\,cos\theta -\left(\tau_{rf}+\tau_{rb}+(F_w+F_g)\times hg+\lambda m \frac{dv}{dt}\times hg\right)\right]\\
        &= \frac{L_a}{L}mg \,cos\theta -\frac{hg}{L}\left[\left(\frac{\mu mgr\,cos\theta}{hg}+F_w+F_g+\lambda m \frac{dv}{dt}\right)\right]\\
        &= \frac{L_a}{L}mg \,cos\theta -\frac{hg}{L}\left[F_f + F_r\left(1- \frac{r_{wheel}}{hg}\right)\right]
    \end{aligned}
\end{equation}
thus we get, 
\begin{equation}
    \begin{aligned}
        F_{t_f max} &= \eta w_f \\
        F_{t_r max} &= \eta w_r
    \end{aligned}
\end{equation}
where $\eta$ is the coefficient of road adhesion.
\subsection{Drive Cycle}
\begin{figure}
     \centering
     \begin{subfigure}[b]{0.48\textwidth}
         \centering
         \includegraphics[width=\textwidth]{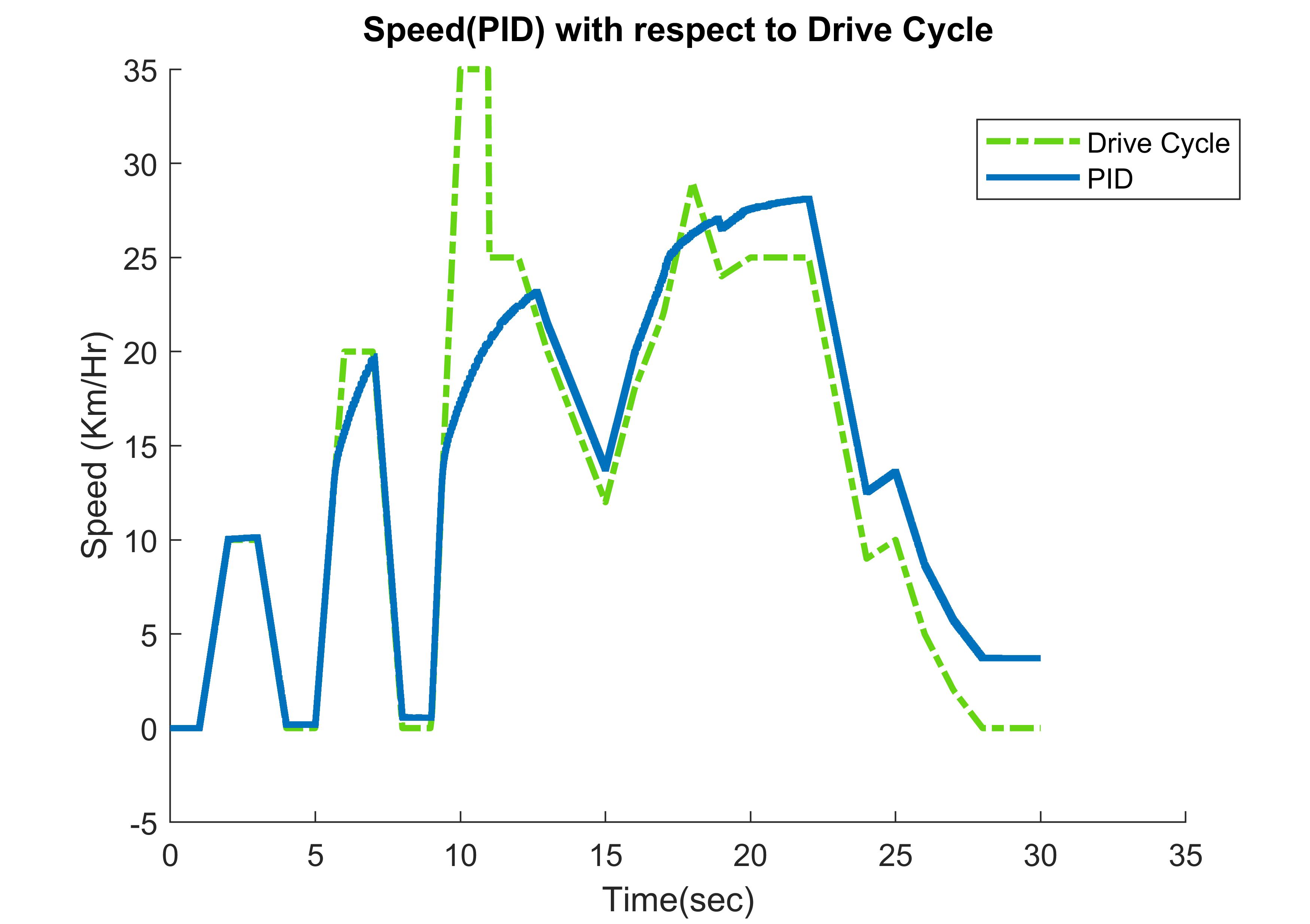}
         \caption{PID control}
         \label{uit_pidspeed}
     \end{subfigure}
     \hfill
     \begin{subfigure}[b]{0.48\textwidth}
         \centering
         \includegraphics[width=\textwidth]{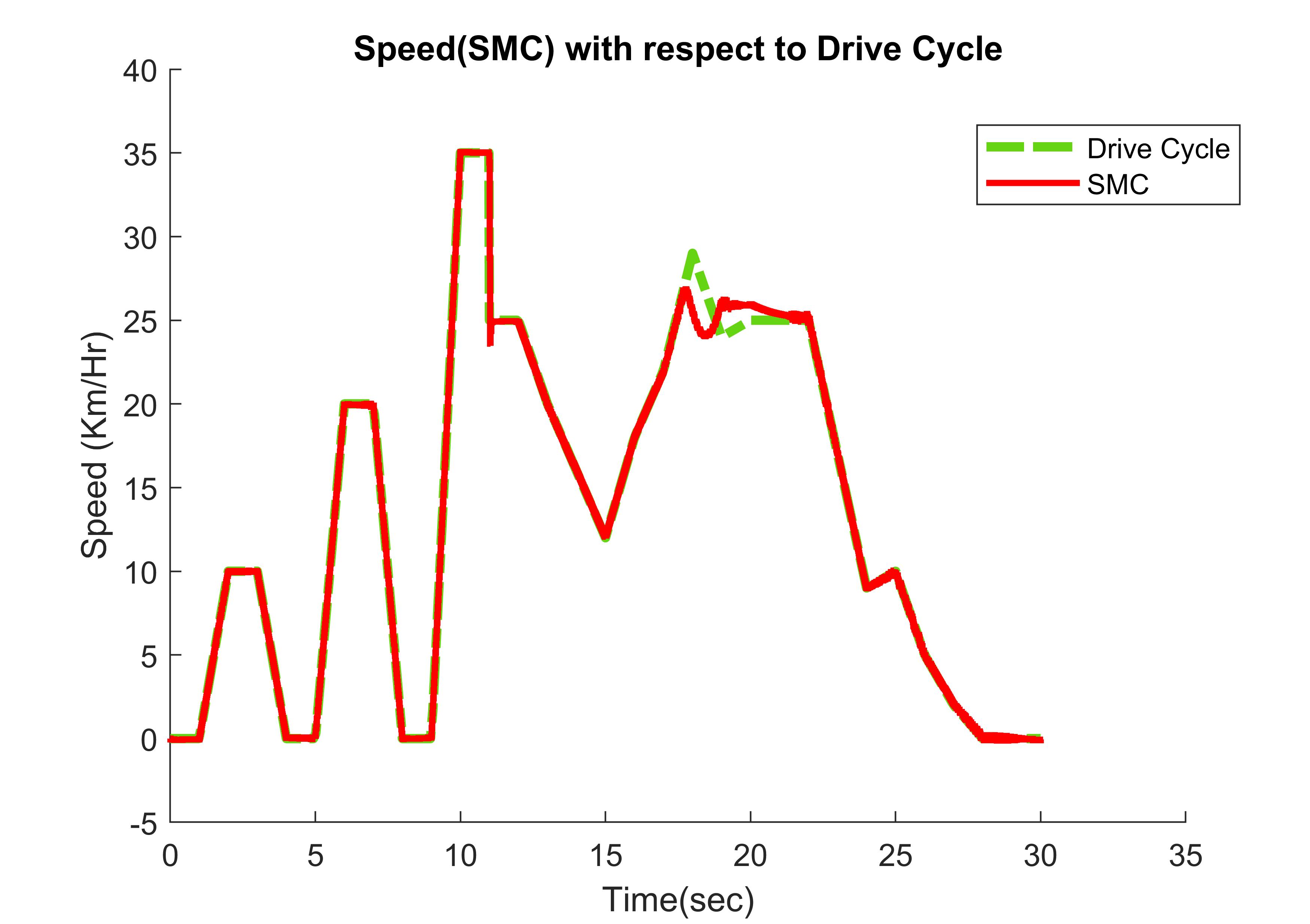}
         \caption{SMC control}
         \label{uit_smcspeed}
     \end{subfigure}
        \caption{Speed Characteristics under the effect of Vehicle Dynamics and Drive Cycle}
        \label{uit_speedcurve}
\end{figure}

The Drive cycle can be used to optimise the inverter's control parameters, reducing losses and improving efficiency by leveraging the power of simulation tools and incorporating real-world driving conditions through the drive cycle, researchers can develop and refine the control strategies involved in the study to improve the overall electric vehicle's performance. This approach allows for the identification of potential issues and opportunities for improvement in the control strategy. The drive cycle represents the trajectory or data of a real world driving scenario and thus is very important for comparative study, performance benchmarking, and other automotive standardisation \cite{schwarzer2012drive}. The drive cycle that has been used has been shown by the green curve in Fig. \ref{uit_speedcurve}.

\par The Vertical axis of the graph shows the vehicle speed (in kmph) and the horizontal axis of the graph shows the time (in seconds) in which the vehicle would have driven. The characteristics of the graph shows that, the nature of driving is quite linear but the later stage shows random driving which implies dynamic behaviour of the driver towards the terrain or system dependent disturbances. The primary intent is to follow the exact drive cycle characteristics by maintaining the other parameters of the drivetrain sound and robust  \cite{froberg2008efficient}.

\subsection{Simulation and Validation}
To test our proposal, three different situations for testing the vehicle drivetrain and control schema have been considered. 
\begin{enumerate}
    \item Open Loop Model: This is the most basic test scenario where only a desired speed is set as a reference without the consideration of a feedback response.
    \item Non-complex Closed Loop Model: Comparative analysis between both the control schemes with respect to the open loop model is done. A closed loop model is essential in understanding how a system reacts to disturbances, thus the interaction of the vehicles' achieved motor speed to desired speed is seen all real-life noises and disturbances are neglected.
    \item Closed Loop Model with Vehicle Dynamics: This segment introduces the concept of Vehicle Dynamics and Drive cycle, which provides a more realistic approach to the real world simulation. In addition to a general closed loop model the various complexities and randomness are taken into consideration. The comparative analysis of the performance of the PID and SMC strategies under this section has been undertaken, the introduction of the vehicle dynamics introduces the real world physical factors while the drive cycle is key in benchmarking data and models.
\end{enumerate}
 \begin{figure}[ht]
    \centering
    \includegraphics[width=0.80\textwidth]{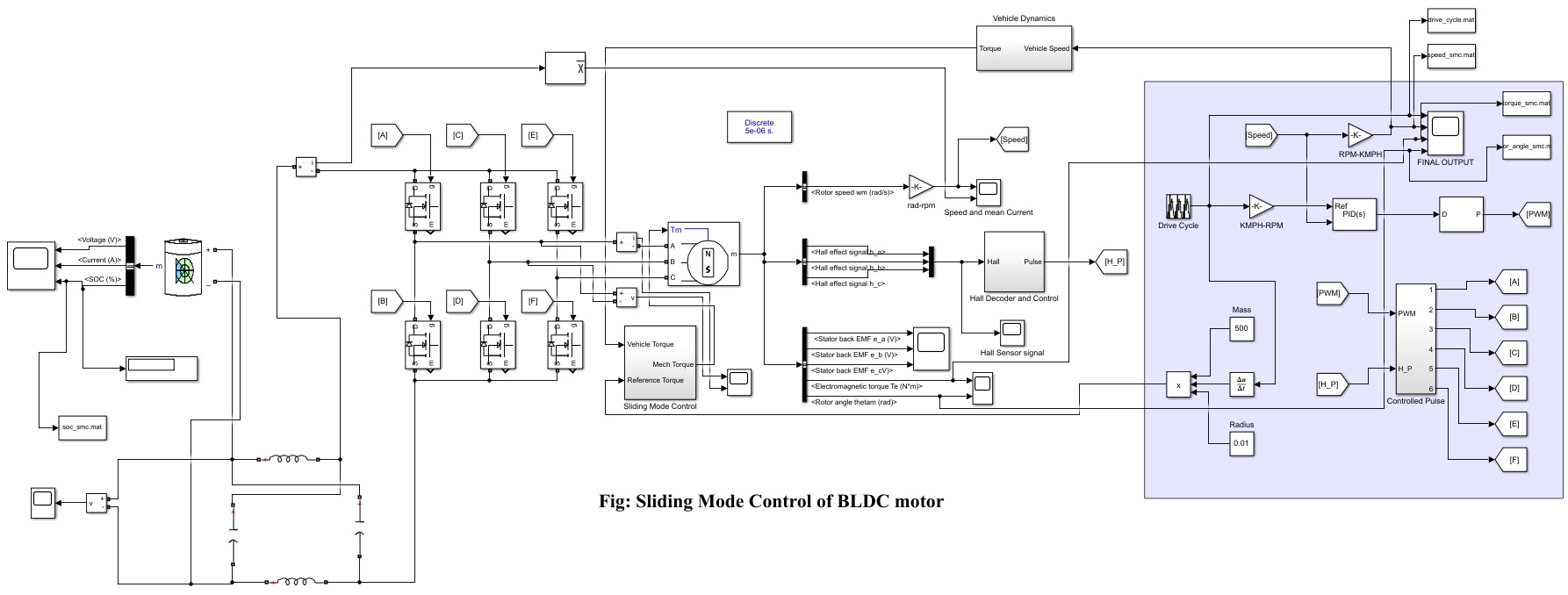}
    \caption{Simulation Design window (MATLAB SIMULINK)}
    \label{uit_simulation}
\end{figure}

The validation of the performance can be done by the computation of Root-Mean-Squared-Error(RMSE). The basic equations for RMSE can be given by:
\begin{equation}
    RMSE = \sqrt{\sum_{i=1}^N \frac{(\chi_i -\hat{\chi_i})^2}{N}}
\end{equation}
where $\chi$ represents the actual signal and $\hat{\chi}$ the estimated signal.

\section{Results and Discussion}
To evaluate the effectiveness of our proposed controller design along with the modified converter topology i.e. ZSI, we compared it with a conventional proportional-integral-derivative (PID) controller in three different scenarios: an open-loop model without feedback, a closed-loop model without randomness, and a closed-loop model with randomness. In each scenario, we measured the speed, torque, rotor angle, and battery state of charge (SoC) of the brushless DC (BLDC) motor.
\subsection{Open Loop Control}
 The first dimension of any research is to set a baseline, and open loop model acts as that, here the motor's reference speed was initially set to 1000 rpm. The graph shows that the motor experiences a primary overshoot close to the reference speed, but due to the lack of controllability, the motor is unable to maintain the desired speed and runs continually to reach the desired speed. Thus the vehicle can never be controlled to the optimum levels. Additionally, the torque curve of the motor also shows that the motor draws a lot of torque and current during the initial stage, the main  point to be noted that this occurs every time there is a change in speed i.e. the reference. Such excessive current can have negative effects on the vehicle's electrical system in the long run.
 \begin{figure}[ht]
    \centering
    \includegraphics[width=0.40\textwidth]{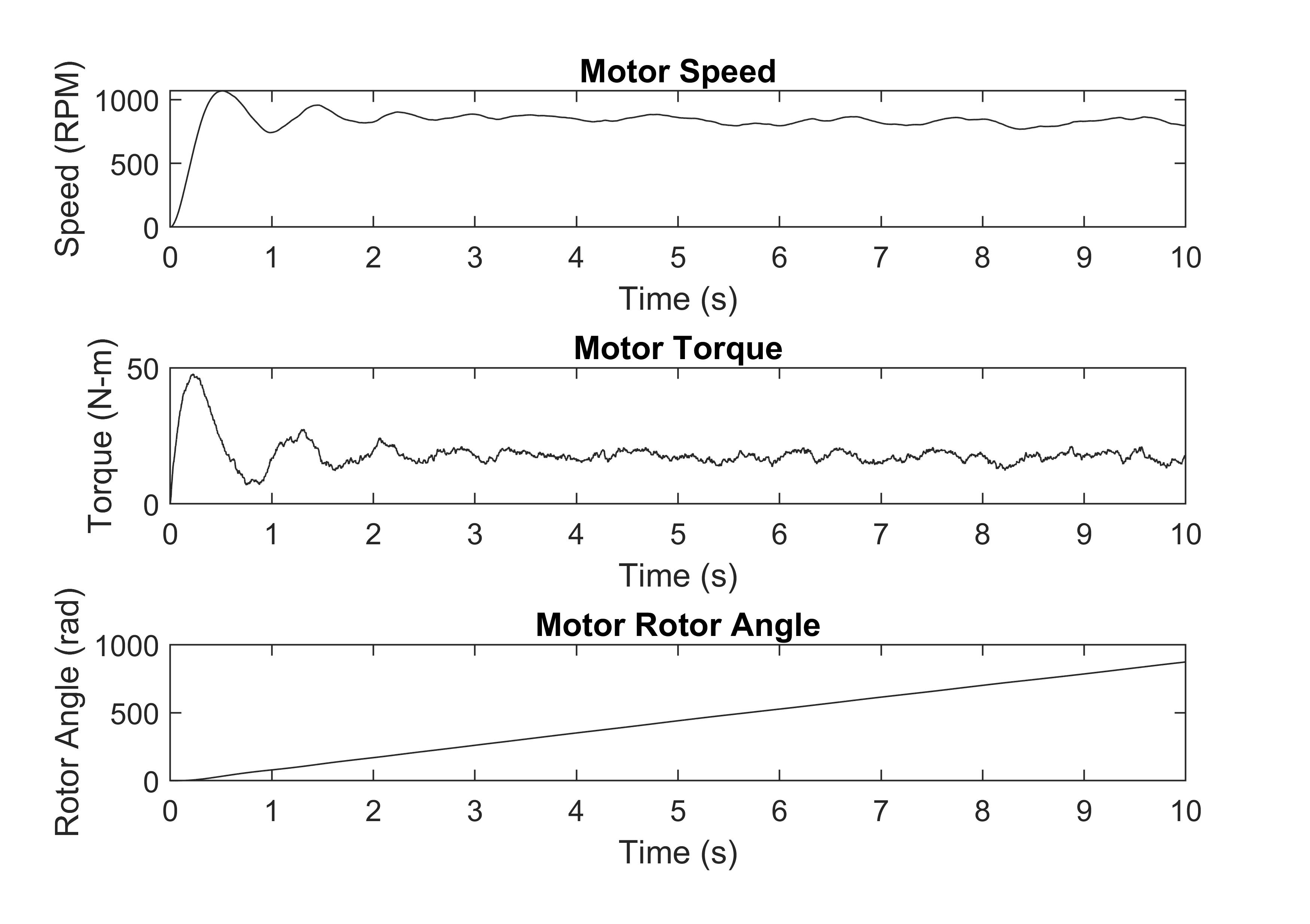}
    \caption{BLDC Performance during Open Loop Control}
    \label{uit_openloop}
\end{figure}
 \subsection{Non-complex Closed Loop Control}
 The closed loop control strategy, through feedback mechanisms, helps eliminate disturbances faced during open loop control from the system and ensures that the motor reaches the desired speed within the required time frame.
In the present segment the BLDC motor's performance has been evaluated using both industrial PID and advanced sliding mode control strategies with a reference speed of 1000 rpm. It is observed that the initial torque is high but it quickly settles down without any randomness. Closed-loop control strategies using both the algorithms demonstrate exceptional performance in controlling the motor's speed, and the motor reaches the desired speed in a short amount of time. In addition, both control strategies effectively remove system noise, leading to stable motor operation. However, it's important to note the absence of complexities introduced by physical phenomena in the process. The working of Closed loop control needs to account for the Vehicle Dynamics as well as consider a drive cycle to compare the performance of the two strategies which will be done in the next segment.
 \begin{figure}[ht]
    \centering
    \includegraphics[width=0.40\textwidth]{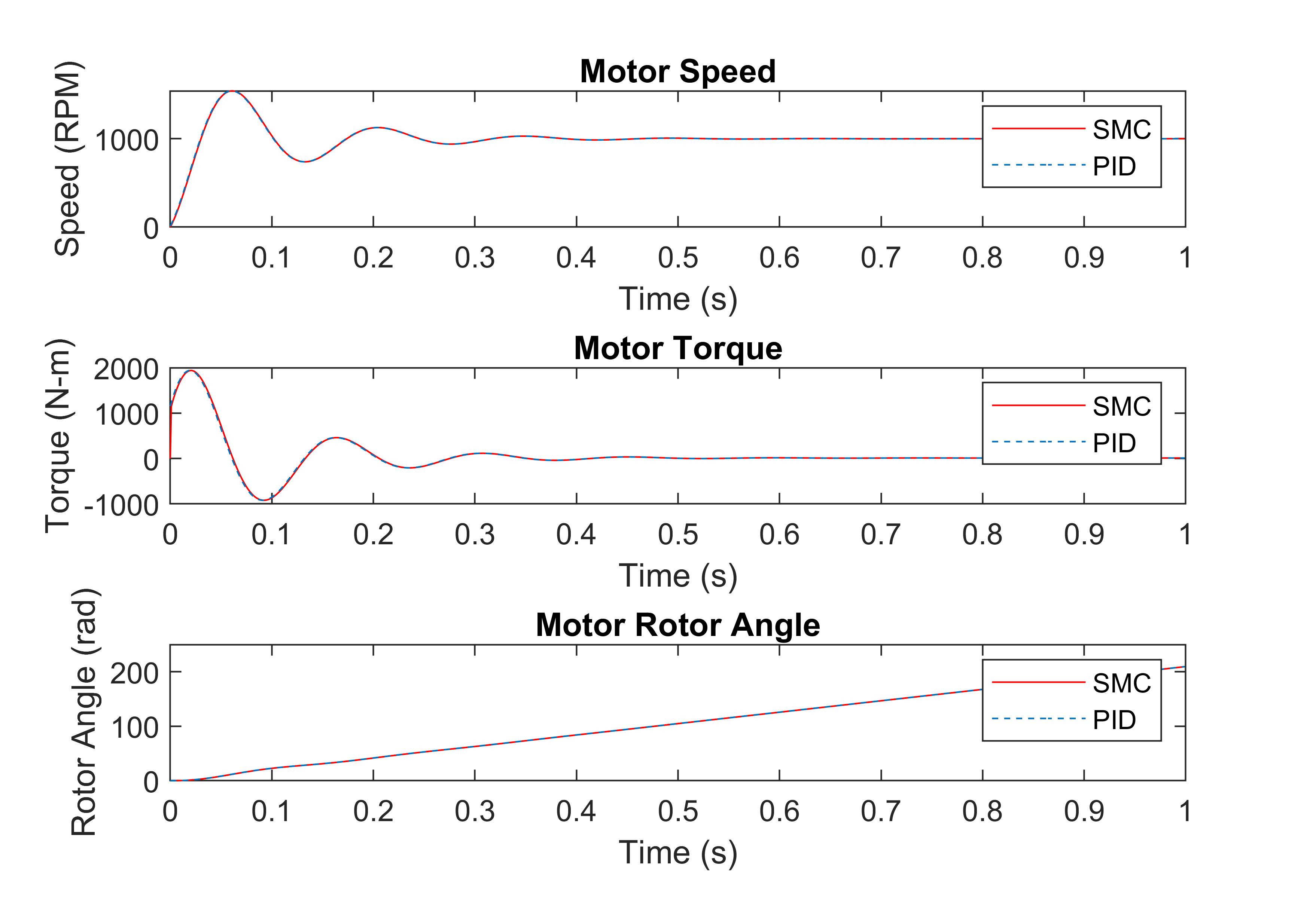}
    \caption{BLDC Performance during Closed Loop Control without the presence of Vehicle Dynamics and Drive Cycle}
    \label{uit_closedloop}
\end{figure}
\subsection{Closed Loop Control with Drive cycle and Vehicle Dynamics}
In order to realistically simulate the driving behaviour of electric vehicles, we incorporated vehicle dynamics and drive cycle blocks into our model. The drive cycle represents a subset of real-world driving environments, where the speed and braking requirements change over time. Our objective entitled designing a controller that could drive the vehicle according to the drive cycle, while maintaining control over the system. Although the torque, SOC, and rotor angle parameters remained consistent between the PID and SMC performances, the controllability of the vehicle remains the primary focus while choosing. \\
\par In general, The study infers that the sliding mode control (SMC) strategy outperforms the PID strategy in terms of speed, torque, and rotor angle tracking, as well as battery SoC management.In the open-loop model, the SMC controller achieved a steady-state speed of $1000$ RPM with a settling time of $0.45$ seconds, whereas the PID controller achieved a steady-state speed of 860 RPM with a settling time of $3.00$ seconds. Similarly, in closed-loop models without randomness, the SMC controller achieved a steady-state torque of $5.6$ Nm with a settling time of $0.45$ seconds, whereas the PID controller achieved a steady-state torque of $17.3$ Nm with a settling time of $3.00$ seconds. The biggest validation metric of the success of SMC is given by RMSE, which is a statistically significant test in error analysis, where SMC recorded $0.5867$ which is exceedingly lower than $4.0140$ of that of PID, denoting better performance of SMC.
\begin{center}
\begin{tabular}{ |p{3cm}||p{3cm}|p{3cm}|p{3cm}|  }
 \hline
 \multicolumn{3}{|c|}{Quantitative Analysis} \\
 \hline
 Parameter Name & PID & SMC\\
 \hline
 RMSE   & 4.0140 & 0.5867 \\ 
 Settling time & 3.00 sec & 0.45 sec \\
 Steady State Speed & 860 RPM & 1000 RPM \\
 \hline
\end{tabular}
\end{center}

\par SMC offers better tracking accuracy, robustness to disturbances and uncertainties, and improved stability over a wide spectrum of operating conditions. Moreover, SMC's capability of achieving higher efficiency and faster response times than traditional control methods, makes it highly desirable option for electric vehicle control applications. Overall, our study highlights the importance of incorporating realism into the simulation of electric vehicle control, and demonstrates the superior performance of SMC over PID control in real-world driving conditions.
\par Furthermore, we found that the SMC strategy was more effective at managing the battery SoC than the PID strategy. Specifically, the SMC controller was able to regenerate the battery SoC within practical capbilities throughout the entire drive cycle, while the PID controller resulted in low SoC dissipation but very feeble or no regeneration. Overall, the results of the study illustrate  superior performance of the SMC strategy over the PID strategy in terms of speed, torque, rotor angle, and battery SoC management for electric vehicle control using for a ZSI.
\begin{figure}
     \centering
     \begin{subfigure}[b]{0.325\textwidth}
         \centering
         \includegraphics[width=\textwidth]{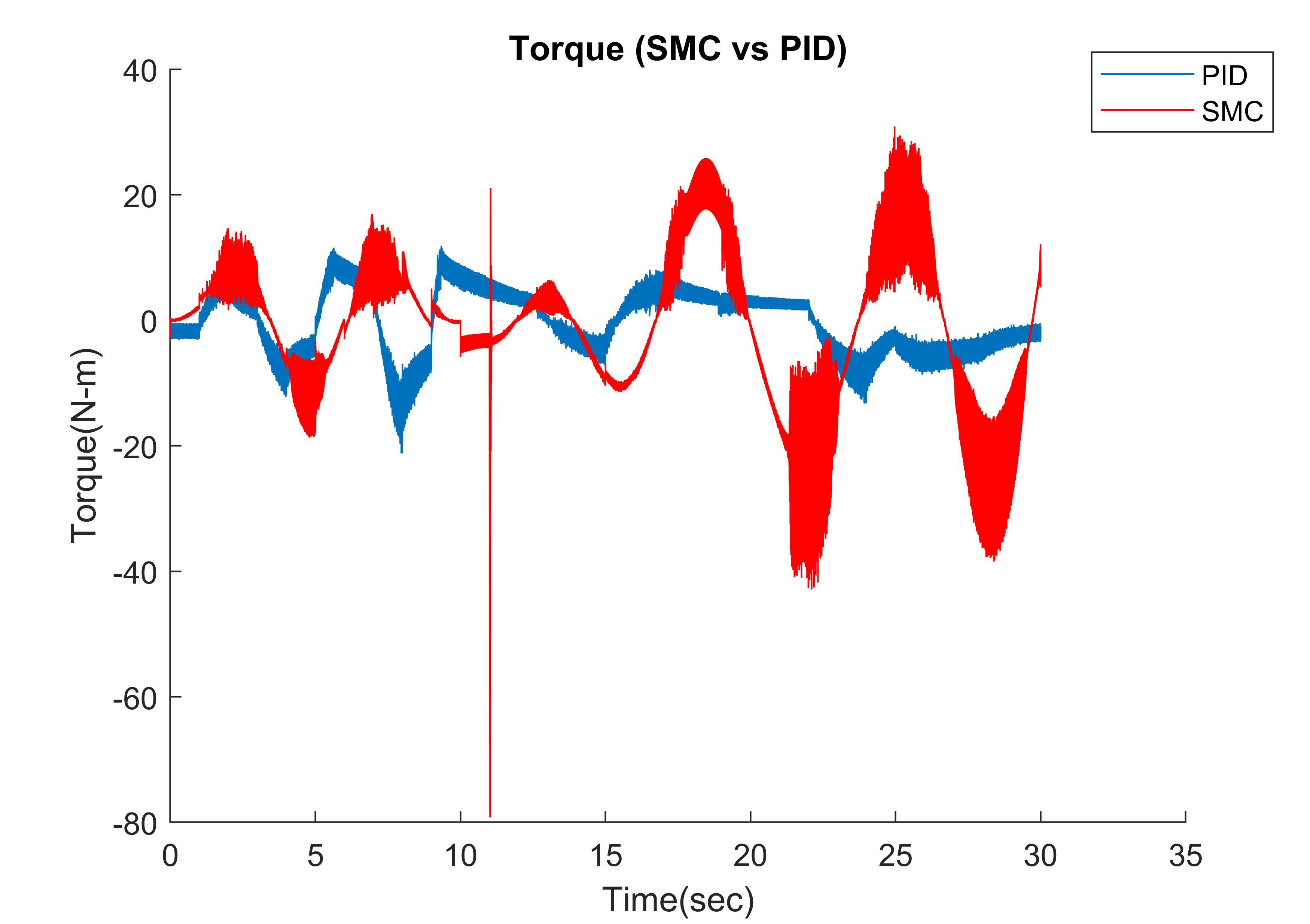}
         \caption{Torque characteristics}
         \label{uit_torque}
     \end{subfigure}
     \hfill
     \begin{subfigure}[b]{0.325\textwidth}
         \centering
         \includegraphics[width=\textwidth]{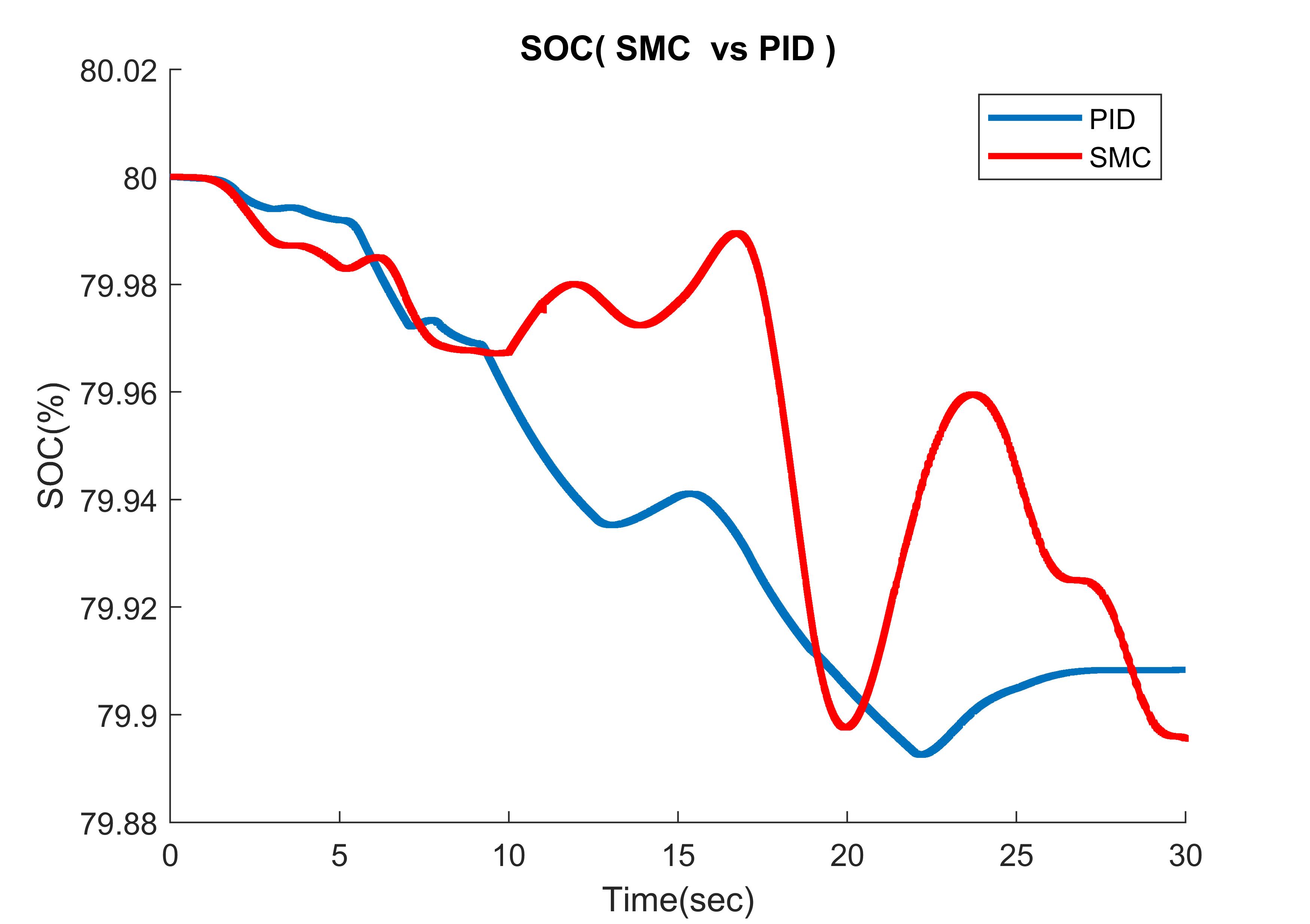}
         \caption{SoC characteristics}
         \label{uit_soc}
     \end{subfigure}
     \hfill
     \begin{subfigure}[b]{0.325\textwidth}
         \centering
         \includegraphics[width=\textwidth]{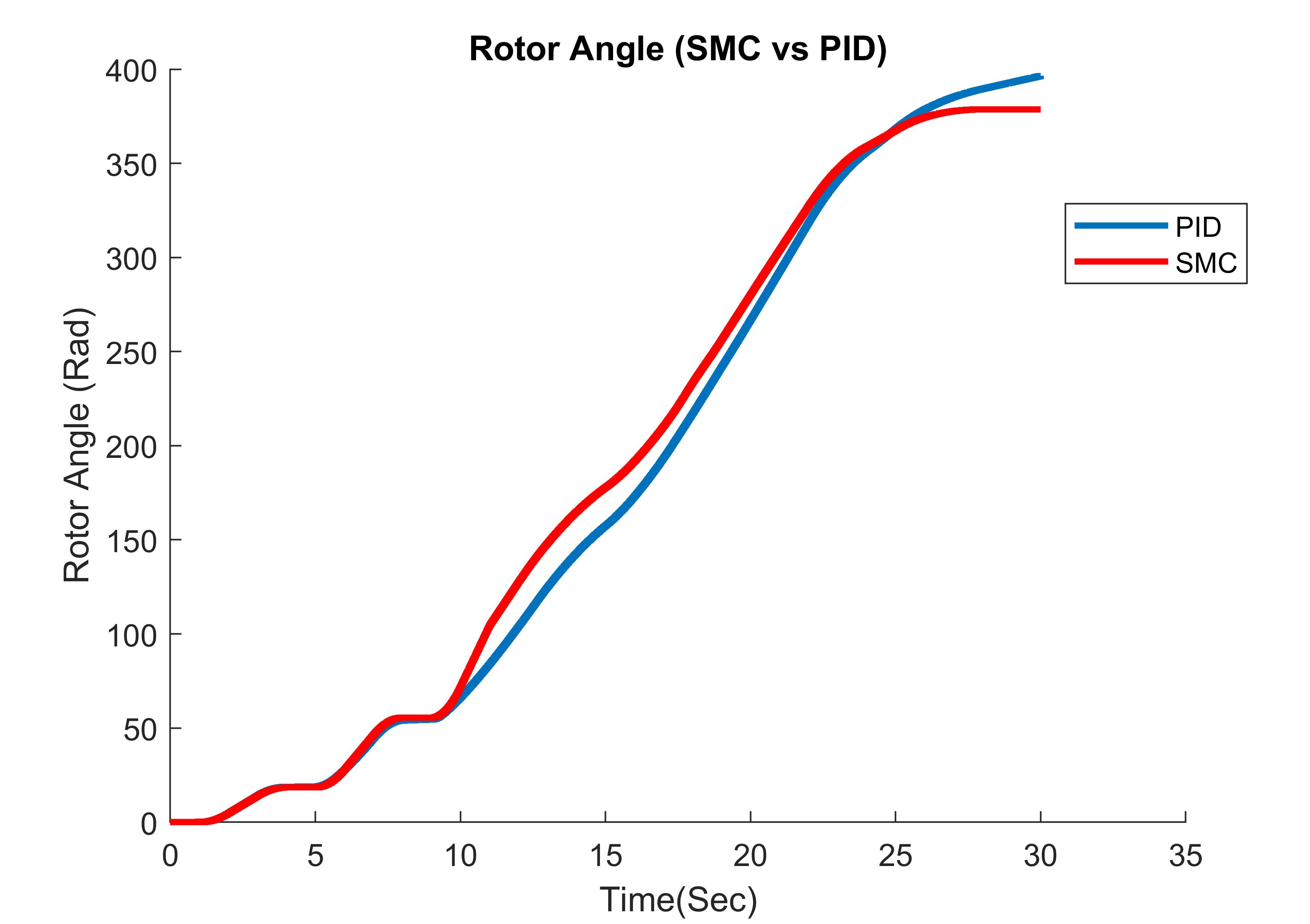}
         \caption{Rotor Angle characteristics}
         \label{uit_angle}
     \end{subfigure}
        \caption{Performance analysis of BLDC motor under Closed loop control with the effect of Vehicle Dynamics and Drive Cycle}
        \label{uit_comparison}
\end{figure}

\section{Conclusion}
The PID control performed well during the first phase of the drive cycle which mostly contained linear driving characteristics, but struggled once the randomness of real-world driving was introduced. Specifically, the PID control could not maintain the desired speed within the required time frame, and could not control the vehicle during fast transitions. However, when we tested the same drive cycle using SMC, we found that it performed exceptionally well. SMC was able to follow the drive cycle and maintain control over the vehicle system, even when faced with the randomness of real-world driving conditions. The RMSE results give us the rigid proof that SMC is performing much better than the usual PID control algorithm. Our results demonstrate that overall the SMC controller outperforms the PID controller and is better suited for controlling electric vehicles in real-world driving environments. The introduction of regenerative braking in substantial form in SMC increases the overall utility and suitability in adoption as this parameter can affect the overall range of the vehicle.
\par This study contributes to the ongoing research efforts to develop more robust and efficient control strategies for electric vehicles, which will ultimately help accelerate the adoption of sustainable transportation technologies. There are possibilities for further research that could provide more viable and employable solutions to the chattering problem that may spawn due to changes in vehicle types and parametrization. Efforts in this direction can accelerate a quicker and more effective development which can be replicated or applied over other electric vehicle drivetrains. In the present scope of research, Hardware-in-loop simulation and hardware implementation using FPGA, could also be performed at further stages to validate the model’s efficiency and applicability in the physical realm.


\begin{thebibliography}{10}

\bibitem{ackermann1998sliding}
J.~Ackermann and V.~Utkin.
\newblock Sliding mode control design based on ackermann's formula.
\newblock {\em IEEE transactions on automatic control}, 43(2):234--237, 1998.

\bibitem{baldursson2005bldc}
S.~Baldursson.
\newblock Bldc motor modelling and control-a
  matlab{\textregistered}/simulink{\textregistered} implementation.
\newblock {\em Chalmers Open Digital Repository}, 2005.

\bibitem{barillas2015comparative}
J.~K. Barillas, J.~Li, C.~G{\"u}nther, and M.~A. Danzer.
\newblock A comparative study and validation of state estimation algorithms for
  li-ion batteries in battery management systems.
\newblock {\em Applied Energy}, 155:455--462, 2015.

\bibitem{borovic2005open}
B.~Borovic, A.~Liu, D.~Popa, H.~Cai, and F.~Lewis.
\newblock Open-loop versus closed-loop control of mems devices: choices and
  issues.
\newblock {\em Journal of Micromechanics and Microengineering}, 15(10):1917,
  2005.

\bibitem{choi2001global}
H.-s. Choi, Y.-h. Park, Y.~Cho, and M.~Lee.
\newblock Global sliding-mode control. improved design for a brushless dc
  motor.
\newblock {\em IEEE control systems magazine}, 21(3):27--35, 2001.

\bibitem{crolla2012impact}
D.~A. Crolla and D.~Cao.
\newblock The impact of hybrid and electric powertrains on vehicle dynamics,
  control systems and energy regeneration.
\newblock {\em Vehicle system dynamics}, 50(sup1):95--109, 2012.

\bibitem{ding2019automotive}
Y.~Ding, Z.~P. Cano, A.~Yu, J.~Lu, and Z.~Chen.
\newblock Automotive li-ion batteries: current status and future perspectives.
\newblock {\em Electrochemical Energy Reviews}, 2:1--28, 2019.

\bibitem{drakunov1992sliding}
S.~V. Drakunov and V.~I. Utkin.
\newblock Sliding mode control in dynamic systems.
\newblock {\em International Journal of Control}, 55(4):1029--1037, 1992.

\bibitem{ehsani2007hybrid}
M.~Ehsani, Y.~Gao, and J.~M. Miller.
\newblock Hybrid electric vehicles: Architecture and motor drives.
\newblock {\em Proceedings of the IEEE}, 95(4):719--728, 2007.

\bibitem{froberg2008efficient}
A.~Froberg and L.~Nielsen.
\newblock Efficient drive cycle simulation.
\newblock {\em IEEE Transactions on Vehicular Technology}, 57(3):1442--1453,
  2008.

\bibitem{hwang2012design}
C.~Hwang, P.~L. Li, C.-T. Liu, and C.~Chen.
\newblock Design and analysis of a brushless dc motor for applications in
  robotics.
\newblock {\em IET electric power applications}, 6(7):385--389, 2012.

\bibitem{jang2002position}
G.~Jang, J.~Park, and J.~Chang.
\newblock Position detection and start-up algorithm of a rotor in a sensorless
  bldc motor utilising inductance variation.
\newblock {\em IEE Proceedings-Electric Power Applications}, 149(2):137--142,
  2002.

\bibitem{jeon2000new}
Y.~Jeon, H.~Mok, G.~Choe, D.~Kim, and J.~Ryu.
\newblock A new simulation model of bldc motor with real back emf waveform.
\newblock In {\em COMPEL 2000. 7th Workshop on Computers in Power Electronics.
  Proceedings (Cat. No. 00TH8535)}, pages 217--220. IEEE, 2000.

\bibitem{jerrelind2021contributions}
J.~Jerrelind, P.~Allen, P.~Gruber, M.~Berg, and L.~Drugge.
\newblock Contributions of vehicle dynamics to the energy efficient operation
  of road and rail vehicles.
\newblock {\em Vehicle System Dynamics}, 59(7):1114--1147, 2021.

\bibitem{johnson2005pid}
M.~A. Johnson and M.~H. Moradi.
\newblock {\em PID control}.
\newblock Springer, 2005.

\bibitem{kaynak2001fusion}
O.~Kaynak, K.~Erbatur, and M.~Ertugnrl.
\newblock The fusion of computationally intelligent methodologies and
  sliding-mode control-a survey.
\newblock {\em IEEE Transactions on Industrial Electronics}, 48(1):4--17, 2001.

\bibitem{levant1993sliding}
A.~Levant.
\newblock Sliding order and sliding accuracy in sliding mode control.
\newblock {\em International journal of control}, 58(6):1247--1263, 1993.

\bibitem{liu2014observer}
J.~Liu, S.~Laghrouche, and M.~Wack.
\newblock Observer-based higher order sliding mode control of power factor in
  three-phase ac/dc converter for hybrid electric vehicle applications.
\newblock {\em International Journal of Control}, 87(6):1117--1130, 2014.

\bibitem{liu2011advanced}
J.~Liu, X.~Wang, J.~Liu, and X.~Wang.
\newblock {\em Advanced sliding mode control}.
\newblock Springer, 2011.

\bibitem{liu2019brief}
K.~Liu, K.~Li, Q.~Peng, and C.~Zhang.
\newblock A brief review on key technologies in the battery management system
  of electric vehicles.
\newblock {\em Frontiers of mechanical engineering}, 14:47--64, 2019.

\bibitem{liu2012permanent}
P.~Liu and H.~Liu.
\newblock Permanent-magnet synchronous motor drive system for electric vehicles
  using bidirectional z-source inverter.
\newblock {\em IET Electrical Systems in Transportation}, 2(4):178--185, 2012.

\bibitem{liu2016impedance}
Y.~Liu, H.~Abu-Rub, B.~Ge, F.~Blaabjerg, O.~Ellabban, and P.~C. Loh.
\newblock {\em Impedance source power electronic converters}.
\newblock John Wiley \& Sons, 2016.

\bibitem{loh2004pulse}
P.~C. Loh, D.~M. Vilathgamuwa, Y.~S. Lai, G.~T. Chua, and Y.~Li.
\newblock Pulse-width modulation of z-source inverters.
\newblock In {\em Conference Record of the 2004 IEEE Industry Applications
  Conference, 2004. 39th IAS Annual Meeting.}, volume~1. IEEE, 2004.

\bibitem{mande2020comprehensive}
D.~Mande, J.~P. Trov{\~a}o, and M.~C. Ta.
\newblock Comprehensive review on main topologies of impedance source inverter
  used in electric vehicle applications.
\newblock {\em World Electric Vehicle Journal}, 11(2):37, 2020.

\bibitem{manzetti2015electric}
S.~Manzetti and F.~Mariasiu.
\newblock Electric vehicle battery technologies: From present state to future
  systems.
\newblock {\em Renewable and Sustainable Energy Reviews}, 51:1004--1012, 2015.

\bibitem{nian2014regenerative}
X.~Nian, F.~Peng, and H.~Zhang.
\newblock Regenerative braking system of electric vehicle driven by brushless
  dc motor.
\newblock {\em IEEE Transactions on Industrial Electronics}, 61(10):5798--5808,
  2014.

\bibitem{park2000new}
S.~J. Park, H.~W. Park, M.~H. Lee, and F.~Harashima.
\newblock A new approach for minimum-torque-ripple maximum-efficiency control
  of bldc motor.
\newblock {\em IEEE Transactions on industrial electronics}, 47(1):109--114,
  2000.

\bibitem{peters2017environmental}
J.~F. Peters, M.~Baumann, B.~Zimmermann, J.~Braun, and M.~Weil.
\newblock The environmental impact of li-ion batteries and the role of key
  parameters--a review.
\newblock {\em Renewable and Sustainable Energy Reviews}, 67:491--506, 2017.

\bibitem{raveendhra2022effects}
D.~Raveendhra, M.~Prashanth, and K.~Sudha.
\newblock Effects of common-mode voltage in zsi-based induction motor drive for
  ev applications.
\newblock In {\em Proceedings of 3rd International Conference on Machine
  Learning, Advances in Computing, Renewable Energy and Communication: MARC
  2021}, pages 385--397. Springer, 2022.

\bibitem{pilehvar2015inverter}
M.~S.~Pilehvar, M.~Mardaneh, and A.~Rajaei.
\newblock An analysis on the main formulas of z-source inverter.
\newblock {\em Scientia Iranica}, 22:1077--1084, 04 2015.

\bibitem{schwarzer2012drive}
V.~Schwarzer and R.~Ghorbani.
\newblock Drive cycle generation for design optimization of electric vehicles.
\newblock {\em IEEE Transactions on Vehicular Technology}, 62(1):89--97, 2012.

\bibitem{shao2006improved}
J.~Shao.
\newblock An improved microcontroller-based sensorless brushless dc (bldc)
  motor drive for automotive applications.
\newblock {\em IEEE Transactions on industry applications}, 42(5):1216--1221,
  2006.

\bibitem{shifat2020effective}
T.~A. Shifat and J.~W. Hur.
\newblock An effective stator fault diagnosis framework of bldc motor based on
  vibration and current signals.
\newblock {\em IEEE Access}, 8:106968--106981, 2020.

\bibitem{shtessel2014sliding}
Y.~Shtessel, C.~Edwards, L.~Fridman, A.~Levant, et~al.
\newblock {\em Sliding mode control and observation}, volume~10.
\newblock Springer, 2014.

\bibitem{singh2019comprehensive}
K.~V. Singh, H.~O. Bansal, and D.~Singh.
\newblock A comprehensive review on hybrid electric vehicles: architectures and
  components.
\newblock {\em Journal of Modern Transportation}, 27:77--107, 2019.

\bibitem{siwakoti2014impedance}
Y.~P. Siwakoti, F.~Z. Peng, F.~Blaabjerg, P.~C. Loh, and G.~E. Town.
\newblock Impedance-source networks for electric power conversion part i: A
  topological review.
\newblock {\em IEEE Transactions on power electronics}, 30(2):699--716, 2014.

\bibitem{sujitha2017res}
N.~Sujitha and S.~Krithiga.
\newblock Res based ev battery charging system: A review.
\newblock {\em Renewable and Sustainable Energy Reviews}, 75:978--988, 2017.

\bibitem{tashakori2012direct}
A.~Tashakori and M.~Ektesabi.
\newblock Direct torque controlled drive train for electric vehicle.
\newblock In {\em Proceedings of the World Congress on Engineering}, volume~2,
  pages 4--6, 2012.

\bibitem{tashakori2013fault}
A.~Tashakori and M.~Ektesabi.
\newblock Fault diagnosis of in-wheel bldc motor drive for electric vehicle
  application.
\newblock In {\em 2013 IEEE Intelligent Vehicles Symposium (IV)}, pages
  925--930. IEEE, 2013.

\bibitem{tashakori2011characteristics}
A.~Tashakori, M.~Ektesabi, and N.~Hosseinzadeh.
\newblock Characteristics of suitable drive train for electric vehicle.
\newblock In {\em International Conference on Instrumentation, Measurement,
  Circuits and Systems (ICIMCS 2011)}, volume~2, pages 51--57. ASME Press,
  2011.

\bibitem{tashakori2011modeling}
A.~Tashakori, M.~Ektesabi, and N.~Hosseinzadeh.
\newblock Modeling of bldc motor with ideal back-emf for automotive
  applications.
\newblock In {\em Proceedings of the World Congress on Engineering}, volume~2,
  pages 6--8, 2011.

\bibitem{thakkar2021electrical}
R.~R. Thakkar.
\newblock Electrical equivalent circuit models of lithium-ion battery.
\newblock {\em Management and Applications of Energy Storage Devices}, 2021.

\bibitem{tibor2011modeling}
B.~Tibor, V.~Fedak, and F.~Durovsk{\`y}.
\newblock Modeling and simulation of the bldc motor in matlab gui.
\newblock In {\em 2011 IEEE International Symposium on Industrial Electronics},
  pages 1403--1407. IEEE, 2011.

\bibitem{yedamale2003brushless}
P.~Yedamale.
\newblock Brushless dc (bldc) motor fundamentals.
\newblock {\em Microchip Technology Inc}, 20(1):3--15, 2003.

\bibitem{young1999control}
K.~D. Young, V.~I. Utkin, and U.~Ozguner.
\newblock A control engineer's guide to sliding mode control.
\newblock {\em IEEE transactions on control systems technology}, 7(3):328--342,
  1999.

\end{thebibliography}
\end{document}